# Synergy of fivefold boost SOT efficiency and field-free magnetization switching with broken inversion symmetry: Toward neuromorphic computing

Badsha Sekh[1], Hasibur Rahaman[1], Subhakanta Das[1], Mitali[2], Ramu Maddu[1], Kesavan Jawahar[1] and S.N. Piramanayagam[1, *]

[1]School of Physical and Mathematical Sciences, Nanyang Technological University, 21 Nanyang Link, 637371, Singapore

[2]Thin Film Laboratory, Department of Physics, Indian Institute of Technology Delhi, Hauz khas, New Delhi, India – 110016

*Corresponding author: prem@ntu.edu.sg

## Abstract

Non-volatile Neuromorphic Computing (NC) elements utilizing Spin Orbit Torque (SOT) provide a viable solution to alleviate the memory wall bottleneck in contemporary computing systems. However, the two challenges, low SOT efficiency ($\xi$) and the need for in plane symmetry breaking field for perpendicular magnetization switching, greatly limit its practical implementation. In this work, the enhanced $\xi$ of Platinum (Pt) SOT layer and field free perpendicular magnetization switching are achieved by integrating thin Ruthenium Oxide ($RuO_2$) layer in our material stack. The optimal $RuO_2$ thickness (0.5 nm) enhances 5.2 times Damping Like (DL) SOT efficiency compared with pure SOT layer (Pt), as determined by hysteresis loop shift measurements, with a relatively low resistivity (90 $\mu\Omega$-cm). Moreover, we achieve 3 times reduction of critical magnetization switching current density compared to reference sample. Our experimental findings also demonstrate Rashba-induced substantial field-free magnetization switching in the presence of an emergent built-in interfacial field. Notably, reliable multi resistance synaptic states are achieved by tailoring the synergistic effects of enhanced SOT and interfacial magnetism. The functionality of synaptic states has been further evaluated by implementing an artificial neural network and achieved image recognition accuracies of approximately 95% and 87% on the MNIST and Fashion-MNIST datasets, respectively. This systematic study paves the way to energy-efficient, field-free SOT synapses for practical NC applications.

## 1. Introduction

Conventional computing architectures, limited by high power consumption and large area requirements, struggle to accommodate recently developed algorithms[1,2] despite significant breakthroughs in Complementary Metal-Oxide-Semiconductor (CMOS)-based frameworks. To overcome these constraints, neuromorphic computing[3–5] is emerging as a research surge, mimicking human brain functionality for efficient and low-power computation. In NC, neurons and synapses function as processing and memory units, respectively. To realize high interconnectivity between neurons[6,7], researchers exploit innovative low power SOT based non-volatile Domain Wall (DW) devices having multiple stable resistance states and thus serve as efficient artificial synapses. These devices rely on the DW position, which require efficient DW pinning[6,7] using complex device structure. However, domain nucleation–induced magnetization reversal results in granular switching[8–10] and provides a viable alternate approach for achieving multiple resistance levels. In the quest for developing next generation spintronic synaptic elements, achieving high-performance spin-current sources remains also a major challenge. In this regard, Spin Hall materials that exhibit a high DL SOT efficiency together with relatively low electrical resistivity ($\rho_{xx}$) are crucial. There are several aspects of increasing the SOT efficiency, such as ion bombardment[11], alloying of heavy metal materials[12], multilayered Heavy Metal (HM) structures[13] etc. However, many of these techniques are not suitable for large-scale production. Therefore, sustained research efforts are required to make SOT-based synaptic devices more efficient and viable for commercial applications. A key strategy pursued in this direction involves exploiting interfacial phenomena to amplify SOT efficiency. As an example, Richa et al. have shown high SOT efficiency due to a proximity magnetic field at $PtSe_2$ interface next to NiFe layer[14]. Despite this, it is not suitable for perpendicular magnetization switching as $PtSe_2$ acquires spin splitting only near NiFe. Along with this, reducing interfacial magnetic anisotropy ($K_S$) energy density is another important factor to achieve higher SOT efficiency. In HM/ferromagnet (FM) systems where SOT is dominated by the spin Hall effect (SHE) of the HM layer, the DL SOT efficiency is primarily determined by two factors: the spin Hall angle ($\theta_{SH}$) of the HM and the spin transparency ($T_{int}$) at the HM/FM interface. Beyond advanced ultra-high-vacuum deposition techniques, improving interfacial properties such as reducing the interfacial spin–orbit coupling parameter $K_S$ can enhance $T_{int}$, and thereby promote more efficient transfer of spin current from the SOT source into the adjacent FM layer. Previous studies have shown that inserting an ultrathin antiferromagnetic oxide layer, such as NiO, can mitigate these effects by modifying interfacial spin scattering[15,16]. Moreover, there is a major concern of achieving perpendicular magnetic anisotropy (PMA) on NiO deposited layer and also issue of intermixing of FM with it while annealing the sample stack to improve the PMA[17]. Furthermore, most of the SOT-based artificial synaptic devices need a symmetry breaking external in plane magnetic field ($H_X$) for perpendicular magnetization switching, which lowers storage density, hinders large-scale integration, and limits scalability. To solve this issue, various alternative symmetry-breaking schemes have been explored, including exchange bias[18], spin–orbit precession torque in magnetic trilayers[19], thickness-gradient engineering[20,21] etc. However, most of these approaches face significant challenges that limit their suitability for industrial-scale fabrication and mass production. To address this challenge, altermagnet $RuO_2$ has garnered considerable interest recently in breaking the symmetry to enable field free switching[22–24]. Nevertheless, the realization of the altermagnetic spin splitting torque requires careful epitaxial growth on a single crystalline substrate, which limits practical integration. In contrast, van der Waals magnet/oxide heterostructures[25] have emerged as an alternative platform, where field-free switching can be achieved through an intrinsic interfacial magnetism. However, their operation at room temperature is limited by low Curie temperature of the ferromagnetic layer. Quite recently, polycrystalline $RuO_2$-based heterostructures[26] have been demonstrated for partial field-free switching. It is highlighted that the observed field free switching requires significant current density with a larger pulse width. Additionally, the use of thick $RuO_2$ layer inevitably may increase the overall device resistance and will dissipate higher energy for superior SOT switching. At the same time, the interfacial spin orbit coupling may reduce the interfacial spin transparency and obstacles in achieving higher SOT efficiency. In addition, the concurrent achievement of high SOT efficiency and field-free perpendicular magnetization switching under energetically favorable conditions have been overlooked. These considerations have motivated the present work.

In this study, using thin RuO$_2$ insertion layer, which is a metallic antiferromagnetic in nature at room temperature[27,28], spin transparency has been improved to enhance charge to spin conversion ratio of Pt as well as we achieve field free perpendicular magnetization switching owing to emergent interfacial effect. Using Pt/RuO$_2$(t)/[Co/Pt]$_{\times 2}$ with optimal RuO$_2$ thickness (t), it has been observed that the damping-like spin–orbit torque efficiency of pure Pt SOT layer increases by 5.2 times, which is several orders higher than reported values[29–31]. We also predict that an out of plane spin polarization ($\sigma_Z$) can be generated at the RuO$_2$ interface, enabling substantial field free switching. Moreover, the polarity reversal of field-free magnetization switching following an initial magnetic field assessment indicates the presence of an intrinsic in-plane interfacial magnetization component. Our experiments present a significant reduction of SOT switching critical current density for optimal $t_{RuO_2}$, establishing Pt/RuO$_2$ combination as a promising spin-current generator offering low switching power, moderate electrical resistivity, and negligible current-shunting. By exploiting distinctive gradual magnetization switching, reliable multilevel resistance states have been realized through the engineered synergy between enhanced SOT efficiency and interfacial magnetism. Furthermore, PyTorch-based testing on MNIST and F-MNIST demonstrates the synaptic functionality and highlights its potential for low-power, high-density neuromorphic hardware.

**Sample preparation and characterization**

For this work, Ta (1 nm) and Pt (5.5 nm) layers have been sequentially deposited on Si/SiO$_2$ substrate by DC magnetron sputtering, serving as the seed and SOT layers, respectively. Subsequently, ultrathin RuO$_2$ films are sputtered at a substrate temperature of 300°C using a mixed gas flow of Ar (12 sccm) and O$_2$ (8 sccm). To suppress intermixing between RuO$_2$ and Co, the remaining multilayers - [Co (0.5)/Pt (0.4)]$_{\times 2}$ /Ru (2) are deposited after cooling the samples to room temperature (as illustrated in Figure 1a). Here the numbers in parenthesis are in nanometer. Ru was chosen as capping layer to protect the sample stack from oxidation. As shown in the $\theta$–$2\theta$ XRD patterns (Figure. 1b), the strong intensity of the Pt (111) crystal planes confirms no degradation of its fcc ordering after RuO$_2$(t) deposition. Additionally, X-ray reflectivity (XRR) measurements (Figure 1c) have been carried out to further evaluate the Pt/RuO$_2$ (t) interface. The presence of well-defined Kiessig fringes[32] in all samples indicate excellent film homogeneity and smooth, well-defined interfaces. The perpendicular magnetic anisotropy of the magnetic layers across different $t_{RuO_2}$ are confirmed by both Kerr hysteresis loops (Figure 1d) and Vibrating Sample Magnetometry (VSM) measurements (supplementary information 1.1). Subsequently, we have fabricated our Hall bar devices with width 5 μm and length 30 μm (Figure 1e) comprising the sample stack using optical lithography followed by ion milling.

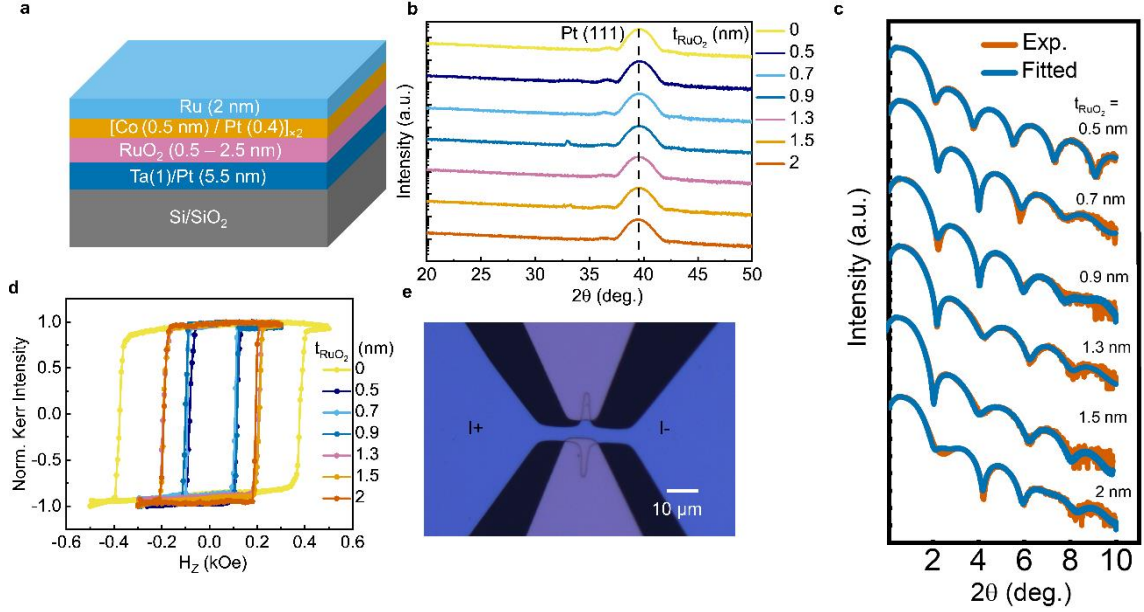

**Figure 1.** Stack configuration, thin film characterization and device fabrication. (a) Stack structure used in this study. RuO₂ thickness has been varied from 0 to 2.5 nm. (b) High resolution X-ray diffraction $\theta$–$2\theta$ scan of the thin film heterostructures (Si/SiO₂/Ta(1)/Pt(5.5)/RuO₂(t)). (c) The experimentally recorded XRR profile for different RuO₂ thicknesses. (c) Measured Kerr hysteresis loops of the thin films in polar mode MOKE microscopy. (d) Optical image of the patterned Hall bar device.

## 2. Enhancement of SOT efficiency with various RuO₂ thicknesses

To estimate the SOT efficiency, we have first performed hysteresis loop shift measurements[33,34] on a reference sample. During the experiment, the writing current (I) is kept constant and $R_{AHE}$ is monitored while sweeping the out of plane field ($H_Z$) at different $H_X$ starting from 0 Oe. The centers of the hysteresis loops are defined as H₀( I$^\pm$)) = [|H$_C^+$ ( I$^\pm$)| - |H$_C^-$ (I$^\pm$)|]/2 and the loop shift, ΔH( I)= H₀( I⁺) - H₀ ( I⁻), quantifies the effective DL SOT field (H$_{DL}$) at a given current I. After that, measurements of ΔH(I) were performed for various $t_{RuO_2}$ samples, along with a control sample where MgO (0.5 nm) was used as the oxide layer instead of RuO₂ (Supplementary Information 1.2). A typical current induced hysteresis loop shift measurement for $t_{RuO_2}$ = 0.5 nm layer sample at H$_X$ = 200 Oe with I$_{DC}$ = 3 mA is shown in Figure.2a (additional details including all measurements data for different $t_{RuO_2}$ in supplementary information 1.3). As depicted in Figure 2.b, H$_{DL}$/I reaches a saturated value when $H_X$ just overcomes the Dzyaloshinskii–Moriya interaction (DMI) effective field |H$_{DMI}$| originated from the Pt/RuO₂/Co interface. Asymmetric charge distribution at the interface of RuO₂/Co may modulate the H$_{DMI}$ as well as H$_{DL}$/I, which varies with different $t_{RuO_2}$ (more details in supplementary information 1.4). Due to various $t_{RuO_2}$ insertion, the effective damping like efficiency[35] per unit current density, $\xi_{DL}^j = \frac{2e}{\hbar} \mu_0 M_s t_{FM} \rho_{xx} H_{DL}/E$, was calculated, where $\hbar$ is the reduced Plank constant, e is the electronic charge, μ₀ is the magnetic permeability, M$_S$ is saturation magnetization (as determined from VSM measurements), $t_{FM}$ is the ferromagnetic layer thickness, $\rho_{xx}$ is the resistivity and E is the electric field. Compared to the reference sample, the lowest $t_{RuO_2}$ exhibits a 5.2-fold enhancement in SOT efficiency, which gradually decreases with increasing its thickness. To further understand the effect, the effective anisotropy energy ($K_{eff}$) was evaluated using the formula[36–38], $K_{eff} = M_S [\int_0^1 H_{OOP} m_{OOP} dm_{OOP} - \int_0^1 H_{IP} m_{IP} dm_{IP}]$, where $H_{OOP}$ and $H_{IP}$ are the applied out of plane and in plane magnetic fields respectively, and $m_{OOP}$ and $m_{IP}$ are the normalized magnetization values in the out-of-plane and in-plane directions, respectively (measurement data has been shown in supplementary information 1.5). The anisotropy field ($\mu_0 H_K$) was calculated as $\mu_o H_k = 2K_{eff}/M_S$. During spin current propagation from the Pt layer into Co, the interfacial spin–orbit coupling (ISOC) competes with the injected spin current. Insertion of a thin $RuO_2$ layer reduces ISOC, which can be indirectly estimated at different $t_{RuO_2}$ via the interfacial magnetic anisotropy energy density ($K_s^{ISOC}$) using the relation[39], $H_K \approx 4\pi M_S - 2K_s^{ISOC}/M_S t$.

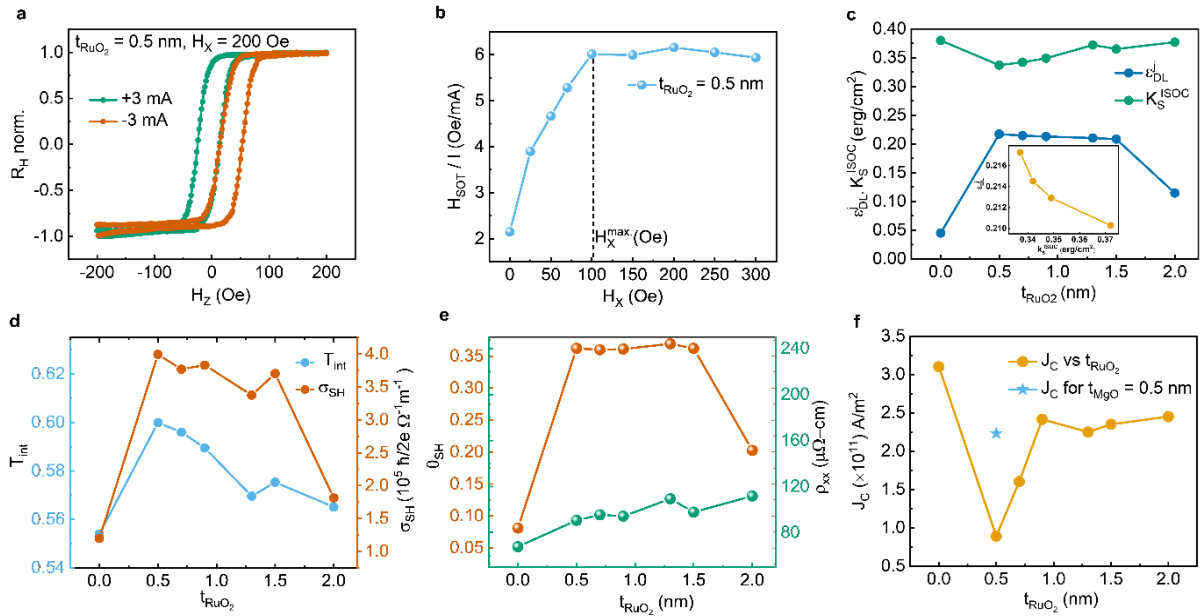

**Figure 2.** Determination of SOT efficiency and its correlation at different $t_{RuO_2}$. (a) Current induced hysteresis loop measurement for $t_{RuO_2}$ = 0.5-layer sample at $H_X$ = 200 Oe with $I_{DC}$ = 3 mA. (b) $H_X$ dependence of SOT induced field per unit applied current ($H_{SOT}/I$). (c) Variation of SOT efficiency per unit current density ($\xi_{DL}^j$) as a function of $t_{RuO_2}$ and relation with interfacial magnetic anisotropy energy ($K_s^{ISOC}$). (d) Estimated interfacial spin transparency ($T_{int}$) and Spin Hall conductivity ($\sigma_{SH}$). (e) Comparison of calculated resistivity ($\rho_{xx}$) and Spin Hall Angle (Θ) with different $t_{RuO_2}$. (f) The variation of critical current density ($J_C$) for SOT driven magnetization switching at different $t_{RuO_2}$.

As shown in Figure.2c, $K_s^{ISOC}$ decreases relative to the reference sample upon thin $RuO_2$ insertion followed by an increasing trend at larger $t_{RuO_2}$. Notably, as $K_s^{ISOC}$ increases, $\xi_{DL}^j$ decreases almost monotonically. At larger $t_{RuO_2}$, however current shunting into the $RuO_2$ layer (as shown the calculated shunting parameters in supplementary information 1.6) also contributes to reduce the SOT efficiency. As described in supplementary information 1.7, the improvement of $\xi_{DL}^E$ of another control sample having $t_{MgO}$ = 0.5 nm ($K_s^{ISOC}$ =0.365 erg/cm², lower than the reference sample) suggests that improvement of the interfacial spin transparency ($T_{int}$) plays a key role in SOT efficiency enhancement. Considering spin backflow (SBF) and spin memory loss (SML) as the primary mechanisms[40] controlling $T_{int}$, the contribution from SBF is given by[41,42] $T_{int}^{SBF}$ = [1-sech ($d_{HM}/\lambda_s$)][1+$G_{HM}$ tanh ($d_{HM}/\lambda_s$)/$2G_{HM/FM}^{\uparrow\downarrow}$]$^{-1}$, where, $\lambda_s$ denotes the spin diffusion length, $G_{HM}$=1/$\rho_{xx}\lambda_s$ represents the spin conductance of the HM layer, and $G_{HM/FM}^{\uparrow\downarrow}$ is the intrinsic spin mixing conductance of the HM/FM interface. For Pt/Co, $G_{Pt} \approx 1.3\times10^{15}$ $\Omega^{-1}$ m$^{-2}$ has been considered from reported thickness-dependent SOT measurements[40]. The theoretical value for Pt/Co interface, $G_{HM/FM}^{\uparrow\downarrow}$ = 0.59×10$^{15}$ $\Omega^{-1}$ m$^{-2}$. SML contributes[41] as, $T_{int}^{SML} = 1 - 0.23 K_S^{ISOC}$ (with $K_S^{ISOC}$ is in erg/cm$^2$), giving the total spin transparency, $T_{int} = T_{int}^{SBF} T_{int}^{SML}$. Using $T_{int}$ parameter, the calculated Spin Hall Conductivity, $\sigma_{SH} = \frac{\hbar}{2e} \xi_{DL}^E / T_{int}$, shows a substantial increase after inserting 0.5 nm RuO$_2$ layer but subsequently decreases at higher $t_{RuO_2}$ (Figure 2d). The Spin Hall Angle ($\theta_{SH}$) is further evaluated as $\theta_{SH} = \frac{\xi_{DL}^j}{T_{int}}$ and is found to be 0.36 for $t_{RuO_2}$ = 0.5 nm, roughly four times higher than that of the reference sample. The insertion of RuO$_2$ slightly increases the resistivity ($\rho_{xx}$) due to its polycrystalline nature and interfacial scattering, but the dependence of $\theta_{SH}$ on $\rho_{xx}$ is not monotonic (Figure. 2e), confirming that the enhancement of $\theta_{SH}$ arises primarily from interfacial spin–orbit modifications. Moreover, we establish the critical switching current density ($J_C$) variation for various $t_{RuO_2}$. $J_C$ is defined as the writing current density at which $R_{xy}$ reaches the maximum switching level under the application of $H_X$ in the reference sample (magnetization switching loops have been shown in Supplementary Information 1.7). For a fair comparison, $J_C$ was evaluated at the same switching percentage and identical $H_X$ for all samples. Under these conditions, we find that $J_C$ is reduced by a factor of three with respect to reference sample upon insertion of the $t_{RuO_2}$ = 0.5 nm layer. As shown in Figure 2f, the trend of $J_C$ with changing $t_{RuO_2}$ is as expected and closely follows its correlation with $\xi_{DL}^j$. However, the insertion of a thin MgO (0.5 nm) layer in the control sample leads to a slight reduction in the critical current density $J_C$, primarily due to the suppression of $K_S^{ISOC}$ and accompanied by an increment in SOT efficiency (supplementary information 1.8). In contrast, the introduction of an ultrathin RuO$_2$ layer results in a pronounced reduction of $J_C$, which can be attributed to the additional contribution from the interfacial Rashba effect together with coherent magnon–mediated[43] spin transport.

### 3. SOT induced field free switching in presence of in-built interfacial field

We first investigate current-induced SOT-driven magnetization switching for the sample with $t_{RuO_2} = 0.5$ nm under various applied $H_X$, starting from 0 Oe, as shown in Figure 3a. Under finite $H_X$, the switching polarity reverses with the field direction, consistent with conventional SOT-driven switching behavior reported previously[44]. Remarkably, robust and reproducible magnetization switching is observed even at $H_X = 0$, demonstrating intrinsic field-free switching in the Pt/RuO$_2$/Co heterostructure. In sharp contrast, no current-induced hysteresis loop is detected in the control sample having MgO (0.5 nm) under zero $H_X$ (see Supplementary Information 2.1), unambiguously highlighting the critical role of the RuO$_2$ insertion layer in enabling field-free magnetization reversal. The switching probability (corresponding MOKE images at different magnetization switching stages have been shown in supplementary information 2.1) increases systematically with increasing current density, reaching a maximum switching value of approximately 68% at the highest applied current density (2.41 × 10$^{11}$ A/m$^2$), as shown in Figure 3b. Here, the switching percentage is defined as $\frac{\Delta R_H}{\Delta R_H^{max}}$, where $\Delta R_H$ is the change in AHE resistance induced by successive current pulses and $\Delta R_H^{max}$ indicates the AHE resistance difference between two opposite magnetization states, as determined from AHE measurement. The repeatability of the field free magnetization switching has been shown in supplementary information 2.2. Zero-field current-induced switching measurements are subsequently extended to all samples with different RuO$_2$ thicknesses, including the reference sample. As shown in Figure 3c, the field-free switching behavior persists only within a finite RuO$_2$ thickness window and vanishes beyond a critical thickness. It should be noted that we have achieved maximum

79% field free switching for $t_{RuO_2} = 0.7$ nm at the same maximum applied current density (2.41 × $10^{11}$ A/m$^2$). To further elucidate the origin of this effect, we have performed hysteresis loop shift measurements at $H_X = 0$ Oe. The extracted loop shift ($\Delta H$) reaches a maximum at $t_{RuO_2} = 0.5$ nm and subsequently decreases with increasing RuO$_2$ thickness, eventually disappearing beyond this value, as shown in Figure 3c. This trend closely correlates with the observed suppression of field-free switching at higher $t_{RuO_2}$. However, the sample with $t_{RuO_2} = 0.5$ nm exhibits a higher spin–orbit torque efficiency and a larger zero-field hysteresis loop shift, whereas more magnetization switching percentage is observed for the sample with $t_{RuO_2} = 0.7$ nm. This apparent discrepancy arises from the fundamentally different roles of torque generation and domain-wall energetics in the magnetization switching process. For $t_{RuO_2} = 0.5$ nm, the system is dominated by strong interfacial effects, giving rise to an enhanced spin–orbit field and consequently, a higher spin–orbit torque efficiency. Moreover, in this ultrathin regime the DMI field remains significantly smaller than the domain-wall anisotropy field (calculation details in the Supplementary Information 2.1), leading to nearly degenerate clockwise and anticlockwise chiral spin configurations and thereby suppressing more effective SOT induced domain-wall propagation. In contrast, for $t_{RuO_2} = 0.7$ nm, the DMI field becomes comparable to the domain-wall anisotropy field, lifting the chiral degeneracy and stabilizing a preferred domain-wall chirality. This stabilization facilitates efficient field-free switching via domain-wall nucleation and propagation, despite a slightly reduced overall SOT efficiency. These findings underscore that deterministic field-free magnetization switching is governed not only by the magnitude of the SOT, but also critically by the balance between the DMI and domain-wall anisotropy, in agreement with previously reported studies[45].

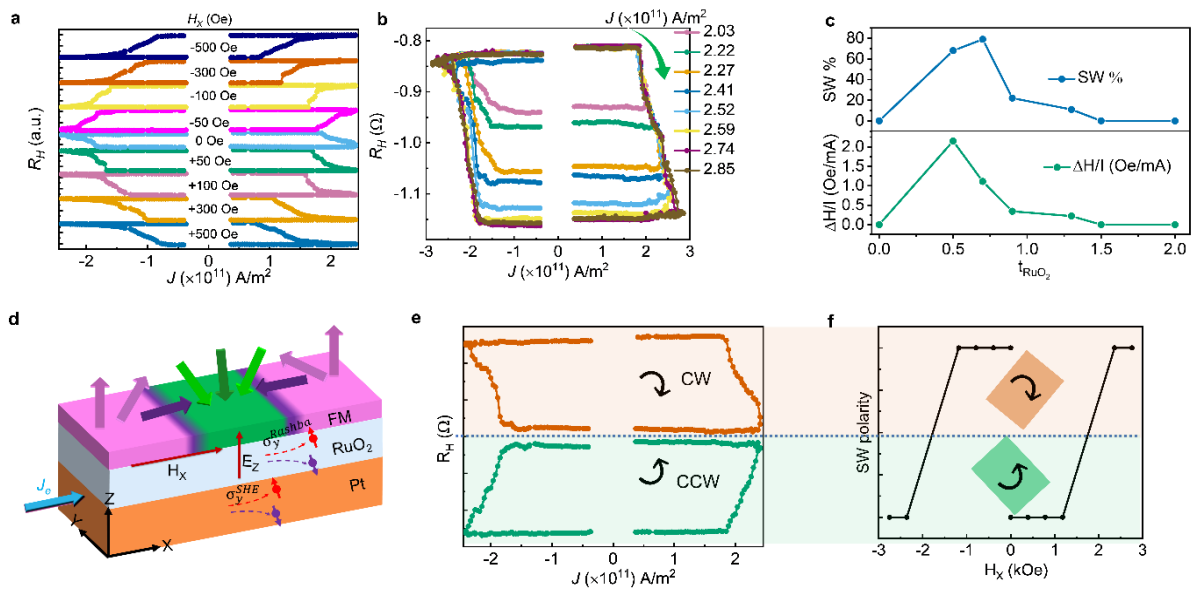

**Figure 3.** SOT induced magnetization switching under presence of symmetry breaking field. (a) SOT switching loops under varying in plane field (H$_X$) starting from zero. (b) SOT induced field free magnetization switching by sweeping various writing currents. (c) The shift of AHE loop ($\Delta H/I$) from loop shift measurement and field free switching percentage (SW %) at different $t_{RuO_2}$. (d) Schematic illustration of the interfacial in-plane field at the Ferromagnet (FM)/oxide interface and the resulting spin polarization from Rashba SOC and the SHE. (e) Clockwise (CW) and counterclockwise (CCW) magnetization reversal observed at zero external magnetic field,

describing deterministic field-free switching. (f) Demonstration of reversing the switching polarity under applied in-plane magnetic fields.

This pronounced thickness dependent field free switching also indicates that ultrathin $RuO_2$ layers generate an unconventional symmetry-breaking torque that enables deterministic field-free switching. In this context, considering a Neel-type domain wall in the Co layer, we propose the following mechanism to explain the observed magnetic field-free switching with varying $RuO_2$ thicknesses (Figure 3d). After applying an in-plane charge current ($J_e$), $\sigma_y$ spin polarization is generated via SHE. Simultaneously, broken inversion symmetry at the $RuO_2$/Co interface gives rise to an interfacial electric field ($E_z$), leading to Rashba-type interfacial spin–orbit coupling which generates $\sigma_y$ interfacial spin accumulation. During the spin angular momentum transfer process to the FM layer, $\sigma_y$ spins experience an effective interfacial in-plane magnetic field ($H_X$), whose magnitude is governed by $RuO_2$(t)/Co interface. This interfacial field couples with the $\sigma_y$ spin and modifies the spin precession, enabling the conversion of in-plane spin polarization into a finite out-of-plane ($\sigma_z$) component. The injected $\sigma_z$ spins subsequently exert a torque on FM layer magnetization, leading to deterministic magnetization switching without any external applied $H_X$. Based on switching polarity measurements, we infer the existence of an interfacial $H_X$ that gives rise to $\sigma_z$ component and the measurement procedure is described as follows. The device is initially saturated with a positive out-of-plane magnetic field and subsequently brought to the zero-field remanent state. Under these conditions, writing current pulses applied at $H_X = 0$, induce a clear clockwise (CW) magnetization switching, as shown in Figure 3e. In contrast, after saturating the device with a negative out-of-plane field and returning to remanence, no field-free switching has been observed. Next, $H_X$ is applied antiparallel to the current, with magnitudes ranging from 390 Oe to 2750 Oe, and is removed prior to each current sweep. If an interfacial $H_X$ exists, a sufficiently strong and oppositely applied $H_X$ can reverse its orientation, leading to field-free switching. Consistent with this expectation, repeated current sweeps reveal a transition to counterclockwise (CCW) magnetization reversal for $H_X \geq 2350$Oe, with the Hall resistance exhibiting the same magnitude change as in positive saturation. The reversal of switching polarity as a function of the magnitude and direction of the applied $H_X$ is schematically illustrated in Figure 3f. Re-saturating the device with a positive out-of-plane field and returning to remanence again produced no field-free switching. Subsequently, $H_X$ has been applied parallel to the current with magnitudes from 390 Oe to 2750 Oe and removed before each sweep. Under these conditions, current sweeps show a CCW-to-CW transition for $H_X \geq 2350$ Oe, revealing a reversal in switching chirality relative to the initial measurement. This behavior provides indirect evidence of the presence of an interfacial $H_X$. The same phenomena have been reproducibly observed in sample with $t_{RuO_2} = 0.7$ nm (Supplementary Information 2.2), confirming its robustness. Importantly, these observations are fully consistent with changes in chirality measured under continuously applied $H_X$, as shown in Figure 3a.

**4. Field free SOT driven Synaptic function and implementation in NC architecture**

In addition to the optimal $J_C$ and highest SOT efficiency observed at $t_{RuO_2}$ = 0.5 nm, it exhibits a broader linear magnetization switching region. These characteristics further make this sample stack suitable to demonstrate the multistate synaptic function using our Hall-bar device. The switching process involves multidomain nucleation, which is ascribed to reduction of effective domain wall anisotropy energy (calculation has been shown in supplementary information 2.2) due to RuO$_2$ insertion. This energy reduction gives rise to multiple intermediate magnetic orientation and is accompanied by SOT induced non-uniform domain-wall motion[46]. To examine the non-volatility and the formation of multi-level magnetization states, we have applied varying pulse current magnitudes ranging from 11 to 15.4 mA (corresponding to the current density ranging from 2.04 × 10$^{11}$ to 2.85 × 10$^{11}$ A/m$^2$) at zero applied external $H_X$. As a result, six distinct resistance states are obtained, closely emulating artificial synaptic behavior (Figure 4a). Upon removal of the writing current after the pulse sequence, the device retains its remanent magnetic state, thereby confirming its non-volatile nature. We have also estimated the power consumption for complete synaptic operation to be about 87.5 μJ. These results underscoring the potential of this SOT-driven device as an energy-efficient NC element. Further reductions in energy consumption can be achieved by device scaling and the use of ultrafast current pulses, which are critical for practical neuromorphic computing.

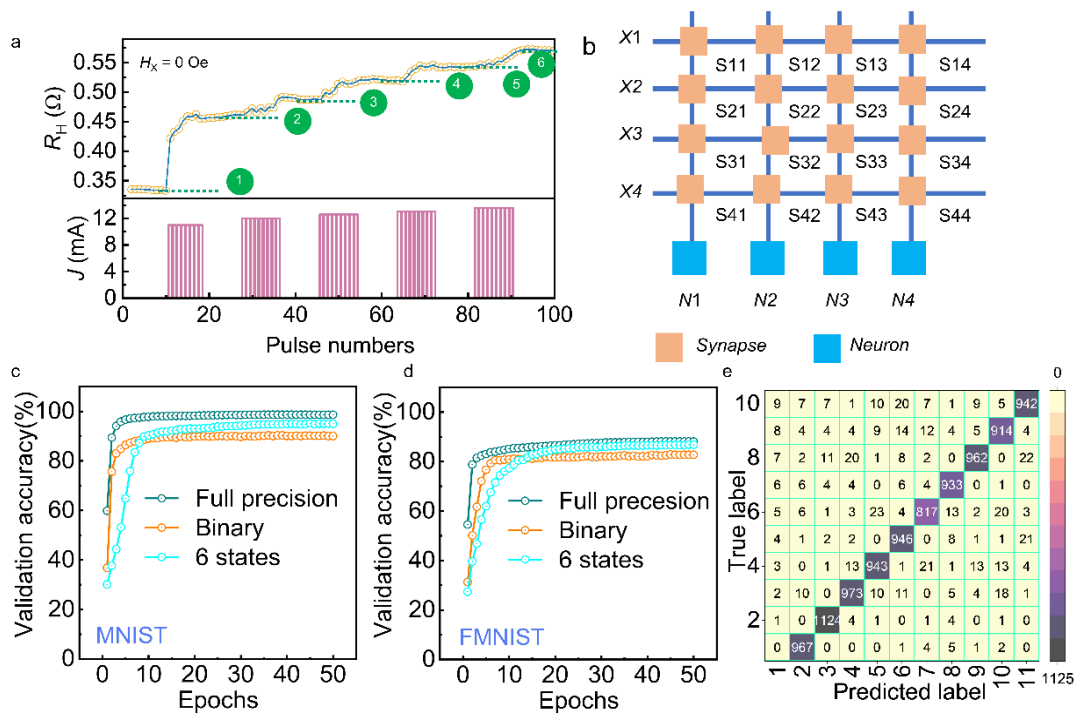

**Figure 4.** Synaptic behavior and ANN assessed for image recognition applications. (a) Multiple AHE resistance states tuned by various pulse train. (b) Schematic illustration showing the integration of Hall-bar devices within a crossbar array. (c) – (d) The dependence of validation accuracy on the number of training epochs for the MNIST and FMNIST datasets is illustrated, respectively. (e) Confusion matrices constructed for the MNIST dataset illustrate the classification outcomes by comparing predicted classes against their corresponding true labels.

Next, the Hall-bar devices are integrated into a crossbar-array architecture to implement the MNIST handwritten digit recognition task, as schematically illustrated in Fig. 4(b)[47,48]. In this framework, the input signal ($X_n$) is encoded as a sequence of current pulses applied to the Hall-bar devices, where each pulse can modulate the synaptic weight, measured through changes in the Hall resistance. The contributions from all synaptic weights within the same column are subsequently accumulated, resulting in a net Hall resistance that represents the weighted sum of the inputs. This accumulated AHE signal is then appropriately scaled and conditioned to serve as the input signal for the subsequent neuron stage. By leveraging the experimentally measured synaptic plasticity characteristics of the devices, the crossbar array can successfully perform the MNIST handwritten digit recognition task. To exploit the NC capability of our SOT devices, artificial neural network (ANN) simulations are carried out using the MNIST and Fashion-MNIST (FMNIST) datasets[49]. In these simulations, the synaptic weight updates are governed by the multi-level resistance states of the fabricated devices, whereas all remaining neuronal computations are implemented digitally using the PyTorch framework. The network performance has been examined under three different precisions configurations: (i) a full-precision floating-point (float32) model serving as the baseline, (ii) a binary (1-bit) weight representation, and (iii) a device-inspired multi-level quantization scheme consisting of six discrete states (approximately 3-bit), directly extracted from experimentally measured resistance values. The multi-state synaptic functionality has been realized by constructing a discrete lookup table (LUT) of normalized weights derived from measured resistance values. Each distinct and stable resistance states (0.33, 0.46, 0.49, 0.52, 0.54, 0.57 $\mu\Omega$) correspond to gradual magnetization switching[50]. These resistance values are subsequently mapped onto the normalized interval [−1, +1], yielding six discrete synaptic weight levels. During network training, continuous weight updates are quantized by projecting them onto the nearest available LUT level, thereby establishing a direct correspondence between experimentally observed resistance states and ANN-level learning.

In the ANN framework two standard image-classification benchmarks are considered: the MNIST dataset, consisting of 70,000 grayscale handwritten digit images (0–9), and the FMNIST dataset, which also contains 70,000 grayscale images representing different categories of clothing items. During training, the validation accuracy is monitored at the end of each epoch to assess learning progress, while the final classification performance is quantified using the independent test set. Figures 4 (c) and 4(d) present the variation of validation accuracy as a function of training epoch for the MNIST and FMNIST datasets respectively. For MNIST, the full-precision network achieves the highest accuracy (98.79%), while the 6-state quantized network closely follows with an accuracy of 95.21% and significantly outperforms the binary case (90.88%). A similar trend is observed for FMNIST, where the 6-state quantized network exhibits performance comparable to the full-precision mode. As expected, MNIST consistently yields higher final accuracies than FMNIST across all quantization schemes due to its relatively simpler visual patterns. These findings highlight the effectiveness and robustness of experimentally derived multi-state synapses across datasets of varying complexity. The corresponding training-loss evolution for all cases is provided in Supplementary Figure S6.

Supplementary Figure S7 compares the final test accuracies across all precision schemes and datasets, confirming that 6-state quantization preserves performance close to the full-precision reference. Figure 4 (e) displays the confusion matrix for the MNIST dataset under the 6-state configuration. Off-diagonal elements are sparse, suggesting only minor confusion between visually similar digit classes. When compared with the confusion matrices obtained for the full-precision and binary networks (Supplementary Figure S8 (b–c)), the 6-state model demonstrates class-wise performance nearly identical to full precision while substantially outperforming the binary case (corresponding confusion matrices for FMNIST are shown in Supplementary Figure S9 (d–f)).

5. Discussion

Spin current generation in the Pt spin–orbit torque (SOT) layer is predominantly governed by its intrinsic spin–orbit interaction. Upon propagation toward the ferromagnetic (FM) layer, however, the efficiency of spin angular momentum transfer becomes highly sensitive to interfacial scattering mediated by interfacial spin–orbit coupling (ISOC). At the lowest RuO$_2$ thickness, the ISOC strength is substantially suppressed, resulting in enhanced interfacial spin transparency. In this regime, the inversion-symmetry-breaking Rashba effect at the Pt/RuO$_2$/Co interface becomes increasingly effective, giving rise to an additional interfacial spin accumulation and a pronounced enhancement of the SOT efficiency. By contrast, insertion of a MgO interlayer leads only to a modest reduction in interfacial magnetic anisotropy and consequently, a comparatively small improvement in SOT efficiency. The significantly larger enhancement has been observed in the Pt/RuO$_2$ system highlights the unique role of RuO$_2$ in engineering interfacial spin transparency beyond what can be achieved through conventional oxide insertion layers. In addition to Rashba-driven interfacial spin generation, the enhanced SOT efficiency at ultrathin RuO$_2$ thicknesses may also involve magnon-mediated spin transport. Specifically, the spin current generated in Pt can excite coherent antiferromagnetic magnons in the RuO$_2$ layer, which subsequently relax by transferring spin angular momentum across the interface over a finite distance. While this mechanism provides a plausible contribution to the observed enhancement, a comprehensive quantitative treatment is beyond the scope of the present study. With increasing RuO$_2$ thickness, the SOT efficiency exhibits a systematic reduction. This behavior can be attributed to strengthened interfacial orbital hybridization between the Co 3d states and the RuO$_2$ 4d–2p hybridized states, leading to an increase in $K_s^{\text{ISOC}}$ and enhanced interfacial spin scattering. At the same time, the higher resistivity of thicker RuO$_2$ layers induces current shunting away from the Pt layer, effectively reducing the spin current injection efficiency and shortening the effective spin diffusion length in Pt. These effects collectively suppress the net SOT efficiency at larger RuO$_2$ thicknesses.

Notably, the enhancement of SOT efficiency and the corresponding reduction in the $J_C$ observed upon insertion of an ultrathin RuO$_2$ layer does not originate from bulk spin Hall effects in RuO$_2$, as its spin Hall angle is considerably smaller[22] than that of Pt. Instead, the dominant contribution arises from interfacial effect due to thin RuO$_2$ layer insertion. Furthermore, when the in-plane spin polarization generated by the combined action of the spin Hall and Rashba effects interacts with an interfacial $H_X$, out of plane spin component, $\sigma_z$ is produced, enabling deterministic field-free magnetization switching. This interfacial magnetic field is likely associated with an in-plane electric potential gradient arising from asymmetric charge distribution, oxygen vacancies in RuO$_2$, or enhanced interfacial magnetism at ultrathin RuO$_2$ thicknesses. As the RuO$_2$ thickness increases, this interfacial interaction progressively weakens and eventually vanishes, consistent with the disappearance of zero-field hysteresis loop shifts. To further corroborate the presence of this interfacial $H_X$, reversible control of the magnetization switching polarity is demonstrated by applying initial $H_X$ prior to current injection, providing indirect experimental evidence for interfacial-field-mediated switching. Collectively, these findings establish RuO$_2$ interfacial engineering as an effective route toward realizing field-free artificial synapses in highly miniaturized neuromorphic devices.

In summary, we have demonstrated 5.2 times enhancement of SOT efficiency compared to reference sample through interfacial engineering of thin RuO$_2$ as an oxide layer. By tuning the interfacial spin interaction, we have achieved a larger charge to spin conversion efficiency (0.36) facilitating 3 times lower critical switching current density. Moreover, we propose and confirm the generation of out of polarization due to interfacial in plane field, which results in substantial field free magnetization switching. Using optimal RuO$_2$ insertion layer, we realize a field-free synaptic device exhibiting six distinct and stable resistance states. The achieved accuracies of 95% on MNIST and 86% on FMNIST demonstrate the great potential of this material engineering in hardware platform for practical neuromorphic computing applications.

## 6. Methodology section

**Sample preparation**: The multilayer thin film heterostructures are deposited on thermally oxidized Si (100) substrates via DC/RF magnetron sputtering. Prior to the sputtering, the deposition chamber is evacuated to a base pressure of $2 \times 10^{-8}$ Torr. The deposition pressure has been maintained at $3 \times 10^{-3}$ Torr during sputtering. After the deposition, thin film stacks have been patterned into Hall bar devices by optical lithography and Ar ion milling. For electrical transport measurements, Ta (5 nm)/Cu (90 nm)/Ta (5 nm) contact electrodes are subsequently deposited by magnetron sputtering.

**Characterization:** The XRD measurements (Gonio scan and XRR) have been performed utilizing a Rigaku SmartLab 3kW HyPix instrument. The magnetic properties measurements of the multilayers are carried out by Magvision MOKE microscope. Magnetization hysteresis loops are performed using a Lake Shore 8600-series vibrating sample magnetometer (VSM). The anomalous Hall effect and current-induced magnetization switching are measured using a Keithley 6221 current source and a Keithley 2182A nanovoltmeter integrated into the MOKE setup.

**Image Recognition:**

To evaluate the pattern-classification capability of the SOT devices, artificial neural network (ANN) simulations have been performed on the MNIST and Fashion-MNIST datasets. The ANN architecture comprises an input layer of 784 neurons corresponding to the flattened $28 \times 28$ grayscale images, followed by two fully connected hidden layers containing 256 and 128 neurons, respectively, and a 10-neuron output layer representing the target classes.

During training, neuron activations are updated deterministically at each forward pass, and synaptic weights have been optimized using gradient-based backpropagation. A cross-entropy loss function is employed to quantify the classification error between the predicted and target labels. The network is trained for 50 epochs using the Adam optimizer for a batch size of 128 with a learning rate of 0.001 and weight-decay regularization to mitigate overfitting. To emulate the hardware constraints, the synaptic weights are quantized into multiple discrete states during training and inference. The resulting classification accuracy and confusion matrices are recorded after each epoch to overall recognition performance on both MNIST and Fashion-MNIST datasets.

## 7. Acknowledgements

The authors gratefully acknowledge the funding from the National Research Foundation (NRF), Singapore for CRP-frontier grant NRF-F-CRP-2024-0012. BS acknowledges the financial assistance from the NTU research scholarship.

## 8. Author contributions

Badsha Sekh: Sample preparation, device fabrication, design of experiments, device testing, analysis, interpretation and writing. Hasibur Rahaman: Device testing, design of experiments, analysis, MNIST work and supporting writing. Subhakanta Das: Device testing and analysis. Mitali: Analysis. Ramu Maddu: Device testing. Kesavan Jawahar: Device testing. S.N. Piramanayagam: Conceptualization of the main idea, design of experiments, analysis, interpretation, and writing.

## 9. Conflict of Interest

The authors declare no conflict of interest.

## 10. Data Availability Statement

The data that support the findings of this study are available from the corresponding author upon reasonable request.

## 11. Keywords

Neuromorphic computing, Large SOT efficiency, Low switching current density, Field free perpendicular magnetization switching.